\documentclass[aps,singlecolumn, showpacs,showkeys,secnumarabic,amssymb,amsmath,superscriptaddress]{revtex4-1}
\usepackage{amssymb,amsmath}
\usepackage{graphics}
\usepackage{graphicx}
\usepackage{bm}
\usepackage{color}
\begin{document}
%
%
%
%

\title{Thermal coherence of the Heisenberg model with Dzyaloshinsky-Moriya interactions in an inhomogenous external field}

\author{Manikandan Parthasarathy}
\affiliation{Department of Physics, 
Ramakrishna Mission Vivekananda College,
Mylapore,
Chennai 600 004, India.}

\author{Segar Jambulingam}
\affiliation{Department of Physics, 
Ramakrishna Mission Vivekananda College,
Mylapore,
Chennai 600 004, India.}

\author{Tim Byrnes}
\affiliation{New York University Shanghai, 1555 Century Avenue, Pudong, Shanghai 200122, China.}
\affiliation{State Key Laboratory of Precision Spectroscopy, School of Physical and Material Sciences, East China Normal University, Shanghai 200062, China}
\affiliation{NYU-ECNU Institute of Physics at NYU Shanghai, 3663 Zhongshan Road North, Shanghai, 200062, China.}
\affiliation{National Institute of Informatics, 2-1-2 Hitotsubashi, Chiyoda-ku, Tokyo 101-8430, Japan}
\affiliation{Department of Physics, New York University, New York, NY 100003, USA}

\author{Chandrashekar Radhakrishnan}
\email{chandrashekar10@gmail.com}
\affiliation{Laboratoire Syst{\`e}mes Complexes et Information Quantify, ESSIE Group, 9 Rue V{\`e}sale, Paris 75005, France}
\affiliation{New York University Shanghai, 1555 Century Avenue, Pudong, Shanghai 200122, China.}
\affiliation{NYU-ECNU Institute of Physics at NYU Shanghai, 3663 Zhongshan Road North, Shanghai, 200062, China.}

\begin{abstract}
The quantum coherence of the two-site XYZ model with Dzyaloshinsky-Moriya (DM) interactions in an external inhomogenous
magnetic field is studied.  The DM interaction, the magnetic field and the measurement basis can be along different directions, and we examine the quantum coherence at finite
temperature.  With respect to the spin-spin interaction parameter, we find that the quantum coherence decreases when the direction of measurement basis is the same as that of the spin-spin interaction.  When the spin-lattice interaction is varied, the coherence always increases irrespective of the relation between its direction and the measurement basis.  Similar analysis of quantum coherence based on the variation of the external inhomogenous magnetic field is also carried out, where we find that the coherence 
decreases when the direction of the measurement basis is the same as that of the external field.  
\end{abstract}

\maketitle

CAPS Number(s): 05.20.G, 05.70.Ce, 02.30.GP 
\section{Introduction}
\label{Intro}
Quantum coherence has been a central concept in the theory of quantum mechanics since its earliest days. In the context of quantum optics, coherence was investigated using 
phase space distributions and higher order correlation functions \cite{glauber1963coherent,sudarshan1963equivalence}. It was only recently that a rigorous quantum information theoretic framework for 
measuring coherence was developed, in the seminal paper by Baumgratz, Cramer and Plenio \cite{baumgratz2014quantifying}.  The work introduced a set of axioms 
which are necessary for a mathematical function to be a proper quantifier of quantum coherence. In Ref. \cite{baumgratz2014quantifying}, 
 two measures 
of quantum coherence were introduced based on these axioms, namely the relative entropy measure of coherence and the 
$\ell_{1}$-norm based quantum coherence. The fundamental properties of quantum coherence have been investigated 
\cite{yao2015quantum,yadin2015quantum,ma2016converting,yadin2016general,bromley2015frozen,du2015conditions,pires2015geometric,
cheng2015complementarity,streltsov2015measuring} and also their applications in different quantum phenomena 
\cite{shi2017coherence,hillery2016coherence,zhang2019demonstrating,ma2019operational,narasimhachar2015low} have been studied.  
These investigations also gave rise to the resource theory framework of quantum coherence
\cite{winter2016operational,brandao2015reversible,chitambar2015relating,chitambar2016critical,del2015resource,
streltsov2015measuring,chitambar2019quantum,radhakrishnan2016distribution}. 

Quantum coherence has been investigated with regards to search algorithms \cite{shi2017coherence,hillery2016coherence}, 
quantum thermodynamics \cite{narasimhachar2015low,lostaglio2015description} and quantum metrology \cite{ma2019operational}.  
Previous studies have established applications of quantum coherence in quantum information processing tasks like quantum dense 
coding and teleportation \cite{pan2017complementarity}.  Another important application of quantum coherence is in the characterization of 
quantum states. Multipartite systems have a large number of degrees of freedom, and it is of interest to understand what type 
of coherences a given state possesses.  Such a scheme was introduced in Ref. \cite{radhakrishnan2016distribution}, 
where coherence could be subdivided into its constituent parts.  Quantum simulation \cite{georgescu2014quantum} is a 
prime example of where this can be applied, since the naturally arising states of a condensed matter system can show 
strikingly different properties depending upon the physical parameters driving phase transitions in the system.  
In the field of condensed matter physics, the measurement and properties of quantum coherence has been studied in various models 
\cite{opanchuk2016quantifying,zheng2016detecting,ishida2013photoluminescence,karpat2014quantum}.  Heisenberg spin models 
\cite{lieb1961two,niemeijer1967some,niemeijer1968some,barouch1970statistical,barouch1971statistical,giamarchi2004quantum} are
one of the important class of models in which coherence has been investigated \cite{malvezzi2016quantum,li2016quantum,cheng2016finite,
ccakmak2015factorization}, where one of the main themes has been the investigation of quantum phase transition\cite{li2016quantum,cheng2016finite,ccakmak2015factorization,radhakrishnan2017quantum2}.
In the Heisenberg XYZ model, the spins have a direct interaction between nearby sites.  
In addition, they can also have a spin-lattice interaction generally referred to as the Dzyaloshinsky-Moriya (DM) interaction
\cite{dzyaloshinsky1958thermodynamic,moriya1960anisotropic}.  Spins are usually aligned parallel or anti-parallel to each other 
depending on whether it is in the ferromagnetic phase or antiferromagnetic phase. 
Due to the spin-lattice interaction the spins are tilted at an angle to the parallel, a feature known as spin-canting.  This spin-canting
usually leads to weak ferromagnetic behavior even in an antiferromagnetic system.  In addition to these inherent properties, 
it is also possible to apply an external magnetic field to the spin system to control and modify its quantum properties.  

A minimal example of the DM model is the two-site model which has been studied to understand the overall dependence of quantities such as entanglement 
\cite{wang2001entanglement,zhou2003enhanced,li2008entanglement,kheirandish2008effect,li2009thermal,gurkan2010entanglement},
quantum discord \cite{guo2011thermal,zidan2014quantum,yi2010thermal}, and coherence  \cite{radhakrishnan2017quantum} as a function 
of the various parameters of the system.  In Ref.  \cite{radhakrishnan2017quantum}, a detailed investigation of quantum 
coherence in two-spin models were carried out.  In this paper, we extend the investigations carried out in Ref.
 \cite{radhakrishnan2017quantum} and investigate two-spin models with DM interactions in the presence of an external field.    
A detailed analysis of the dependence of the quantum coherence on various 
parameters such as the spin-spin interaction, the spin-orbit interaction, DM interaction,  and external magnetic field
 is carried out.  As quantum coherence is a basis dependent quantity, in our work we choose 
different orientations of the DM interaction parameter and the external fields keeping the measurement direction in the $\sigma_{z}$ 
basis for all cases.  This will help us to understand the basis-dependent nature of coherence with respect to the internal parameters such as
spin-spin interaction, as well as the DM interaction parameter.  Finally, the relation between quantum coherence and the external driving 
forces such as the homogeneous component and the inhomogenous component of the magnetic field will be explored.  
 
The article is structured as follows:  In Sec. \ref{samedirection} we discuss the quantum coherence dependence in the two-spin XYZ model 
with DM interactions and external field in the same direction.  The features of quantum coherence where the two-site 
XYZ model with the DM interaction and external field in different directions is discussed in Sec.  \ref{differentdirection}.  Sec. \ref{two control parameters}
gives the simultaneous variation of the quantum coherence with spin-spin interaction and any one of the other parameters
such as DM interaction, homogeneous component, or the inhomogenous component of the external field.  Finally, we present 
our conclusions in Sec. \ref{conclusions}.

%
%
%
\section{The Heisenberg model with DM interaction and the coherence measure}
\label{modeland measure}
The Heisenberg spin model consists of a collection of interacting spins arranged periodically on a lattice.  
The Hamiltonian of the Heisenberg spin chain is written as
\begin{equation}
H = \sum_{n} J_{x} \sigma_{n}^{x}  \sigma_{n+1}^{x} + J_{y} \sigma_{n}^{y}  \sigma_{n+1}^{y} + J_{z} \sigma_{n}^{z}  \sigma_{n+1}^{z},
\label{heisenbergchain}
\end{equation}
where $\sigma_{n}^{x,y,z}$ are the Pauli spin matrices at the site $n$ and $J_{x,y,z}$ are the spin-spin interaction terms in the 
$x$,$y$ and $z$ directions.   For $J_{x} \neq J_{y} \neq J_{z}$, 
the spin model is referred to as XYZ model. The model is antiferromagnetic when $J_{i} > 0$ $(i=x,y,x)$ and ferromagnetic for $J_{i} < 0$.  Along 
with the spin-spin interaction we may also have a spin-lattice interaction which are known as the Dzyaloshinsky-Moriya (DM) 
interaction.  The Hamiltonian of the XYZ spin chain with the DM interaction is 
\begin{equation}
H  = \sum_{n} J_{x} \sigma_{n}^{x} \sigma_{n+1}^{x}  + J_{y} \sigma_{n}^{y} \sigma_{n+1}^{y} +  J_{z} \sigma_{n}^{z} \sigma_{n+1}^{z} 
       + \vec{D} . (\vec{\sigma}_{n} \times \vec{\sigma}_{n+1}) 
\label{xyzhamiltonian}       
\end{equation}
 The antisymmetric spin-lattice interaction 
is accounted for by the DM interaction coefficient $\vec{D}$. 

In our work we further consider an inhomogenous external magnetic field.
The Hamiltonian of a two-spin system with DM interaction 
in an external inhomogenous field is 
\begin{equation}
H =  J_{x} \sigma_{1}^{x}  \sigma_{2}^{x} + J_{y} \sigma_{1}^{y} 
 \sigma_{2}^{y} + J_{z} \sigma_{1}^{z}  \sigma_{2}^{z}
       + \vec{D}  \cdot (\vec{\sigma}_{1} \times \vec{\sigma}_{2}) +  (\vec{B} + \vec{b}) \cdot \vec{\sigma}_{1}^{j} 
       +  (\vec{B} - \vec{b}) \cdot \vec{\sigma}_{2}^{j}.
\label{xyzhamiltonianwithdmandfield}        
\end{equation}
 This field can be written as two fields of the form $\vec{B} + \vec{b}$
acting on the first qubit and the field $\vec{B} - \vec{b}$ acting on the second qubit, where $B$ is the homogeneous and $b$ 
is the inhomogenous component of the external magnetic field.   

From the Hamiltonian of this model, the quantum properties can be studied by examining its eigenstates.  In our work, we investigate the quantum coherence of the system at finite (non-zero) temperature.  Quantum coherence is measured as the distance between  the density matrix $\rho$ and  the decohered density matrix $\rho_{d}$. The decohered density matrix is defined according to the matrix elements
\begin{equation}
\langle \sigma_{1} \sigma_{2} |  \rho_{d} | \sigma_{1}^{\prime} \sigma_{2}^{\prime} \rangle
 =  \langle \sigma_{1} \sigma_{2} |  \rho | \sigma_{1}^{\prime} \sigma_{2}^{\prime} \rangle \delta_{\sigma_{1},\sigma_{1}^{\prime} }
 \delta_{\sigma_{2},\sigma_{2}^{\prime}} .  
\end{equation}
In Ref. \citep{baumgratz2014quantifying}, two different measures of coherence namely the relative 
entropy measure belonging to the entropic class and the $\ell_{1}$-norm of coherence which is a geometric measure were 
introduced.  Another measure based on the quantum version of the Jensen-Shannon divergence 
\cite{lin1991divergence,lamberti2008metric,briet2009properties} was introduced in Ref. \cite{radhakrishnan2016distribution} 
\begin{equation}
\mathcal{D} (\rho,\sigma) \equiv \sqrt{J(\rho,\sigma)} =  \sqrt{   S \left( \frac{\rho + \sigma}{2}  \right) - \frac{S(\rho)}{2} - \frac{S(\sigma)}{2}  }.
\end{equation}
Here $\rho, \sigma $ are arbitrary density matrices and $S = - {\rm Tr} \rho \log \rho$ is the von Neumann entropy.  
This measure has the advantage that it has 
both geometric and entropic properties and is also a metric satisfying triangle inequality. The coherence is then defined in our case as
\begin{equation}
{\cal C} (\rho) = \mathcal{D} (\rho,\rho_d) .
\label{qjsdmeasure}
\end{equation}
The coherence of the two-spin XYZ model in Eq. (\ref{xyzhamiltonianwithdmandfield}) at finite temperature  is discussed in the rest of the paper.   Since quantum coherence is a basis-dependent quantity 
we examine the spin model in two different cases namely:  ($i$) when the control parameters and measurement basis are in the same 
direction; ($ii$) when the control parameters like DM interaction and the external field are orthogonal to the measurement basis.

\section{DM interaction and the external magnetic field along the same direction}
\label{samedirection}

\subsection{$\vec{D} \propto \vec{x}, \vec{B} \propto \vec{x} $ case}
\label{xdirectiondmandfield}
For the case where both the DM interaction and the external magnetic field are in the 
$x$ direction, the Hamiltonian reads 
\begin{equation}
H = J_{x} \sigma_{1}^{x} \sigma_{2}^{x} + J_{y} \sigma_{1}^{y} \sigma_{2}^{y} 
    + J_{z} \sigma_{1}^{z} \sigma_{2}^{z}+D_{x}(\sigma_{1}^{y} \sigma_{2}^{z}-\sigma_{1}^{z} \sigma_{2}^{y})
    +(B_x + b_x ) \sigma_{1}^{x}+(B_x - b_x ) \sigma_{2}^{x},
\label{xyz_ham_inhom_DxBx}     
\end{equation}
where $D_{x}$ is the DM interaction in the $x$-direction, and $B_{x}$ is the average magnetic field and $b_{x}$ is the inhomogneous
magnetic field in the $x$-direction.  The matrix representation of the Hamiltonian in the  $\sigma_{z}$-basis is
\begin{equation}
H = \left(
\begin{array}{cccc}
J_{z}&G_2&G_3&J_{-}\\
G_4&-J_{z}&J_{+}&G_1\\
G_1&J_{+}&-J_{z}&G_4\\
J_{-}&G_3&G_2&J_{z}\\
\end{array}
\right),
\label{ham_mat_xyz_inhom_DxBx}
\end{equation}
where $G_{1,2} = iD_{x}+B_{x} \pm b_x$, $G_{3,4} = G_{1,2}^{*}$ and $J_{\pm} = J_{x} \pm J_{y}$.  The eigenvalues of the Hamiltonian are 
\begin{equation}
E_{1,2}= J_{x} \pm \omega_{1}, \qquad 
E_{3,4}= -J_{x} \pm \omega_{2},
\label{eval_xyz_DxBx}
\end{equation}
and their corresponding eigenvectors are 
\begin{eqnarray}
|\psi_{1,2} \rangle &=& \frac{1}{\sqrt{2}}(\sin \varphi_{1,2}|00\rangle + \cos \varphi_{1,2}|01\rangle+ \cos \varphi_{1,2}|10\rangle + \sin \varphi_{1,2}|11\rangle) \\
|\psi_{3,4} \rangle &=& \frac{1}{\sqrt{2}}(\sin \varphi_{3,4}|00\rangle + \chi \cos \varphi_{3,4}|01\rangle - \chi \cos \varphi_{3,4}|10\rangle - \sin \varphi_{3,4}|11\rangle) \\ \nonumber 
\end{eqnarray}
where the various factors used are 
\begin{eqnarray}
\chi &=&  \frac{-iD_{x}-b_{x}}{\sqrt{b_{x}^{2}+D_{x}^{2}}} ,   \quad
                 \omega_{1} = \sqrt{4B_{x}^{2}+(J_{y}-J_{z})^{2}},  \quad
                \omega_{2} = \sqrt{4b_{x}^{2}+4D_{x}^{2}+(J_{y}+J_{z})^{2}} \nonumber \\
\varphi_{1,2} &=& \arctan \left(\frac{2B_{x}}{J_{y}-J_{z}\pm \omega_{1}}\right),   \quad
\varphi_{3,4}=\arctan \left(\frac{2\sqrt{b_{x}^{2}+D_{x}^{2}} }{-J_{y}-J_{z}\pm \omega_{2}}\right).       
\end{eqnarray}

The state of the two-spin spin system at thermal equilibrium is given by the thermal density matrix $\rho(T) = \exp(- \beta H) /Z$ where
$Z = {\hbox {Tr}} [\exp(-\beta H)]$ is the partition function of the system and $\beta = 1/k_{B} T$, and $k_{B}$ and $T$ are the Boltzmann constant 
and temperature respectively.  For the sake of convenience we assume $k_{B} =1$ throughout our discussion.  The matrix form of the thermal density 
matrix in the $\sigma_{z}$-basis is 
\begin{equation}
\rho(T) = \left(
\begin{array}{cccc}
u_1&q_{1}^{*}&q_{2}^{*}&u_2\\
q_{1}&v_{1}&v_{2}&q_{2}\\
q_{2}&v_{2}&v_{1}&q_{1}\\
u_2&q_{2}^{*}&q_{1}^{*}&u_{1}\\
\end{array}
\right).
\label{tdm_xyz_DxBx}
\end{equation}
The elements of the thermal density matrix (\ref{tdm_xyz_DxBx}) are
\begin{eqnarray}
u_{1,2} &=& \frac{1}{2Z} \left[e^{-\frac{J_{x}+\omega_{1}}{T}} \sin^{2}  \varphi_{1}+ e^{-\frac{J_{x}-\omega_{1}}{T}} \sin^{2}  \varphi_{2}\pm e^{\frac{J_{x}-\omega_{2}}{T}} \sin^{2}  \varphi_{3} \pm e^{\frac{J_{x}+\omega_{2}}{T}} \sin^{2}  \varphi_{4} \right], \\
v_{1,2} &=& \frac{1}{2Z} \left[e^{-\frac{J_{x}+\omega_{1}}{T}} \cos^{2}  \varphi_{1}+ e^{-\frac{J_{x}-\omega_{1}}{T}} \cos^{2}  \varphi_{2}\pm e^{\frac{J_{x}-\omega_{2}}{T}} \cos^{2}  \varphi_{3} \pm e^{\frac{J_{x}+\omega_{2}}{T}} \cos^{2}  \varphi_{4} \right], \\ 
q_{1,2} &=& \frac{1}{2Z} \Big[e^{-\frac{J_{x}+\omega_{1}}{T}} \sin\varphi_{1} \cos\varphi_{1}+ e^{-\frac{J_{x}-\omega_{1}}{T}} \sin\varphi_{2} \cos\varphi_{2} 
                    \pm e^{\frac{J_{x}-\omega_{2}}{T}} \chi \sin\varphi_{3} \cos\varphi_{3}  \nonumber  \\
            &  & \pm e^{\frac{J_{x}+\omega_{2}}{T}} \chi \sin\varphi_{4} \cos\varphi_{4} \Big].
\end{eqnarray}
where the partition function is
\begin{equation}
Z = 2 \left[ e^{-\frac{J_{x}}{T}} \cosh \left(\frac{\omega_{1}}{T}\right) +e^{\frac{J_{x}}{T}} \cosh \left(\frac{\omega_{2}}{T}\right)\right].
\end{equation}
From the density matrix $\rho(T)$, we can write down the decohered density matrix $\rho_{d}$  using which we can measure coherence 
introduced through Eq. (\ref{qjsdmeasure}).   

\begin{figure}
\includegraphics[scale=1.25]{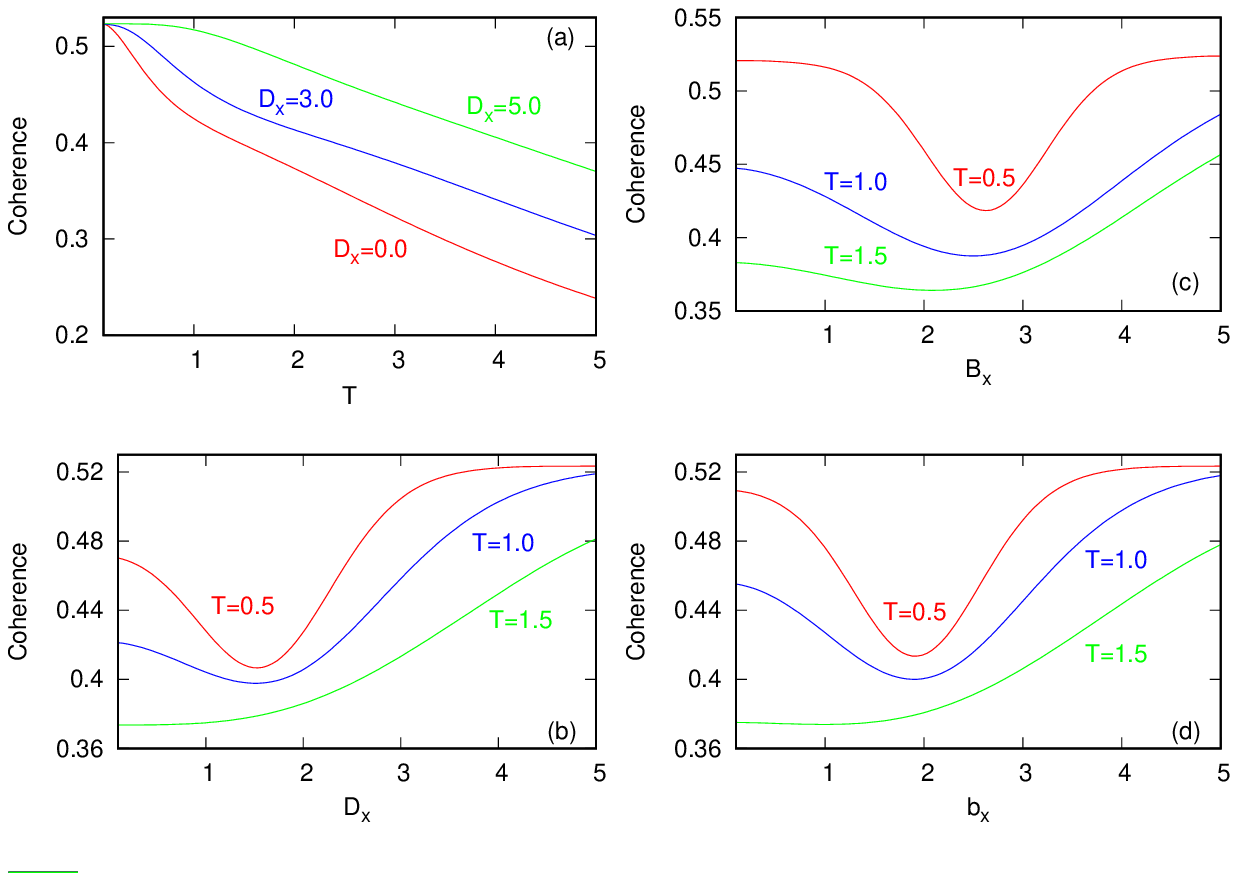}
\caption{Quantum coherence measurement in $\sigma_{z}$ basis in the XYZ model with DM interactions in the $x$-direction and the external 
magnetic field in the $x$-direction.  (a) Quantum coherence versus temperature $T$ for different $D_{x}$, for $B_{x}=3.0$, $b_{x} =1.5$, 
$J_{x}=0.8$, $J_{y}=0.5$ and $J_{z}=0.2$.  (b) Quantum coherence versus $D_{x}$ for different temperatures $T$, for 
$B_{x}=3.0$, $b_{x} =1.5$, $J_{x}=0.8$, $J_{y}=0.5$ and $J_{z}=0.2$.  (c) Quantum coherence versus $B_{x}$ for different temperatures $T$, 
for $D_{x}=1.0$, $b_{x} =1.5$, $J_{x}=0.8$, $J_{y}=0.5$ and $J_{z}=0.2$. (d) Quantum coherence versus $b_{x}$ for different temperatures 
$T$, for $D_{x}=1.0$, $B_{x}=3.0$,  $J_{x}=0.8$, $J_{y}=0.5$ and $J_{z}=0.2$. }
\label{fig1}
\end{figure}

The behavior of quantum coherence is shown in Fig. \ref{fig1}.  Since quantum coherence is a basis dependent quantity, the choice 
of measurement basis plays an important role.  In the Hamiltonian of Eq. (\ref{xyz_ham_inhom_DxBx}), the DM interaction and the field are in the 
$x$-direction, and the measurement is carried out in the $z$-direction.  This behavior of coherence is very similar to the entanglement results 
reported in Ref. \cite{li2008entanglement}.  In Fig. \ref{fig1}(a) the variation of coherence with temperature is shown and we find that coherence decreases with 
temperature.  This can be explained by the thermal fluctuations inducing decoherence in the quantum system, thereby decreasing coherence.
The rate of decrease of coherence is lower for higher values of the DM interaction parameter.  This is because the DM interaction creates an extra
spin-lattice coupling apart from the existing spin-spin coupling.  This additional spin lattice coupling gives rise to additional coherence in the quantum 
system.  The variation of quantum coherence with the DM interaction parameter is shown in Fig. \ref{fig1}(b) for different values of 
temperature.  We find that for lower values of temperature ($T=0.5, T=1.0$), the coherence decreases initially, reaches a minimum value and 
then increases to reach a saturation value.  Initially at $D_{x}=0$, the spin-spin interaction is dominant and when it is increased the spin-lattice 
interaction is slowly introduced.  The initial decrease in the spin-lattice interaction might be due to the competing effects of the 
spin-spin and spin-lattice interaction.  The increase later on can be attributed to the co-operation between the spin-spin and spin-lattice interaction.  
At higher values of temperature ($T=1.5$), the coherence increases gradually to a saturation value. 

The variation of the quantum coherence is investigated as a function of the external magnetic field in Fig. \ref{fig1}(c) 
and \ref{fig1}(d).  We now examine the role played by the external inhomogenous field.  In Fig.  \ref{fig1}(c)  we show the variation of 
quantum coherence with the field $B_{x}$ and we find that for the lower values of temperature, the coherence initially 
decreases and reaches a minimum value and then increases to attain a saturation value.  For the higher values of temperature, 
the coherence increases with the homogeneous field $B_{x}$. The overall behavior of the quantum coherence with 
the inhomogenous field $b_{x}$ resembles the variation of coherence with the field $B_{x}$ as we can see from a comparison between Figs.
\ref{fig1}(c) and \ref{fig1}(d).  On comparison with the earlier works Ref. \cite{li2008entanglement} we find that quantum coherence 
exhibits behavior very similar to entanglement.  From concurrence measurements one can see that entanglement also 
increases with the DM interaction.  The main reason is that the entire coherence comes from the two-spin correlation in a 
manner similar to entanglement.

\subsection{ $\vec{D} \propto \vec{z}, \vec{B} \propto \vec{z} $ case  }
\label{zdirectiondmandfield}
For the case where the DM interaction and the external magnetic field along the $z$-direction, the Hamiltonian reads 
\begin{equation}
H = J_{x} \sigma_{1}^{x} \sigma_{2}^{x} + J_{y} \sigma_{1}^{y} \sigma_{2}^{y} 
    + J_{z} \sigma_{1}^{z} \sigma_{2}^{z}+D_{z}(\sigma_{1}^{x} \sigma_{2}^{y}-\sigma_{1}^{y} \sigma_{2}^{x})
    +(B_z + b_z ) \sigma_{1}^{z}+(B_z - b_z ) \sigma_{2}^{z}.
\label{xyz_ham_DzBz}     
\end{equation}
Here, $B_{z}$ is the average external magnetic field in the $z$ direction and $b_z$ is the degree of inhomogenity  of the field in the $z$-direction.  In the standard
measurement basis ($\sigma_{z}$ basis) the Hamiltonian can be written in matrix form as 
\begin{equation}
H = \left(
\begin{array}{cccc}
2B_{z}+J_{z}&0&0&J_{x}-J_{y}\\
0&2b_{z}-J_{z}&2iD_{z}+J_{x}+J_{y}&0\\
0&-2iD_{z}+J_{x}+J_{y}&-2b_{z}-J_{z}&0\\
J_{x}-J_{y}&0&0&-2B_{z}+J_{z}\\
\end{array}
\right).
\label{ham_mat_xyz_DzBz}
\end{equation}
The eigenvalues and eigenvectors corresponding to the Hamiltonian are 
\begin{eqnarray}
E_{1,2} &=& J_{z} + \omega_{1} \qquad   |\psi_{1,2} \rangle = \sin \theta_{1,2} |00 \rangle + \cos \theta_{1,2} |11 \rangle,  \\
E_{3,4} &=& -J_{z} + \omega_{2} \qquad   |\psi_{3,4} \rangle = \sin \theta_{3,4} |01 \rangle + \chi  \cos \theta_{3,4} |10 \rangle,  
\end{eqnarray}
where the factors $\omega_{1}$ and $\omega_{2}$ are
\begin{equation} 
\omega_{1} = \sqrt{4B_{z}^{2}+(J_{x}-J_{y})^{2}},  \qquad  \omega_{2} = \sqrt{4b_{z}^{2}+4D_{z}^{2}+(J_{x}+J_{y})^{2}}.
\end{equation}
In the expression for the eigenvectors the parameters used are
\begin{equation}
\theta_{1,2}=\arctan \left(\frac{J_{x}-J_{y}}{\pm \omega_{1}-2B_{z}}\right),  \qquad
\theta_{3,4}=\arctan \left(\frac{\sqrt{(J_{x}+J_{y})^{2}+D_{z}^{2}}}{\pm \omega_{2}-2b_{z}}\right)
\end{equation}
and  $\chi =  \frac{J_{x}+J_{y}-2iD_{z}}{\sqrt{(J_{x}+J_{y})^{2}+4D_{z}^{2}}}$.
The thermal density matrix $\rho(T)$ in the $\sigma_{z}$-basis is
\begin{equation}
\rho(T) =\frac{1}{2Z} \left(
\begin{array}{cccc}
m_{1}&0&0&m_{3}\\
0&n_{1}&n_{3}&0\\
0&n_{4}&n_{2}&0\\
m_{3}&0&0&m_{2}\\
\end{array}
\right).
\label{tdm_xyz_DzBz}
\end{equation}
The matrix elements of the thermal density matrix (\ref{tdm_xyz_DzBz}) are
\begin{eqnarray}
m_{1,2} &=&  \mp e^{\frac{-J_z}{T}} \left[\cosh \left(\frac{\mu}{T}\right) \mp \frac{B_z}{\mu} \sinh \left(\frac{\mu}{T}\right)  \right] \\
m_{3} &=& - e^{\frac{-J_z}{T}} \left[\frac{J_{-}}{\mu} \sinh \left(\frac{\mu}{T}\right)\right] \\ 
n_{1,2} &=& e^{\frac{J_z}{T}} \left[\cosh \left(\frac{\nu}{T}\right) \mp \frac{b_z}{\nu} \sinh \left(\frac{\nu}{T}\right)  \right] \\
n_{3,4} &=& - e^{\frac{-J_z}{T}} \left[\frac{J_{+} \pm iD_{z}}{\nu} \sinh \left(\frac{\nu}{T}\right)\right]             
\end{eqnarray}
where the partition function 
\begin{equation}
Z = 2 \left[ e^{-\frac{J_{z}}{T}} \cosh \left(\frac{\mu}{T}\right) +e^{\frac{J_{z}}{T}} \cosh \left(\frac{\nu}{T}\right)\right].
\end{equation}

\begin{figure}
\includegraphics[scale=1.25]{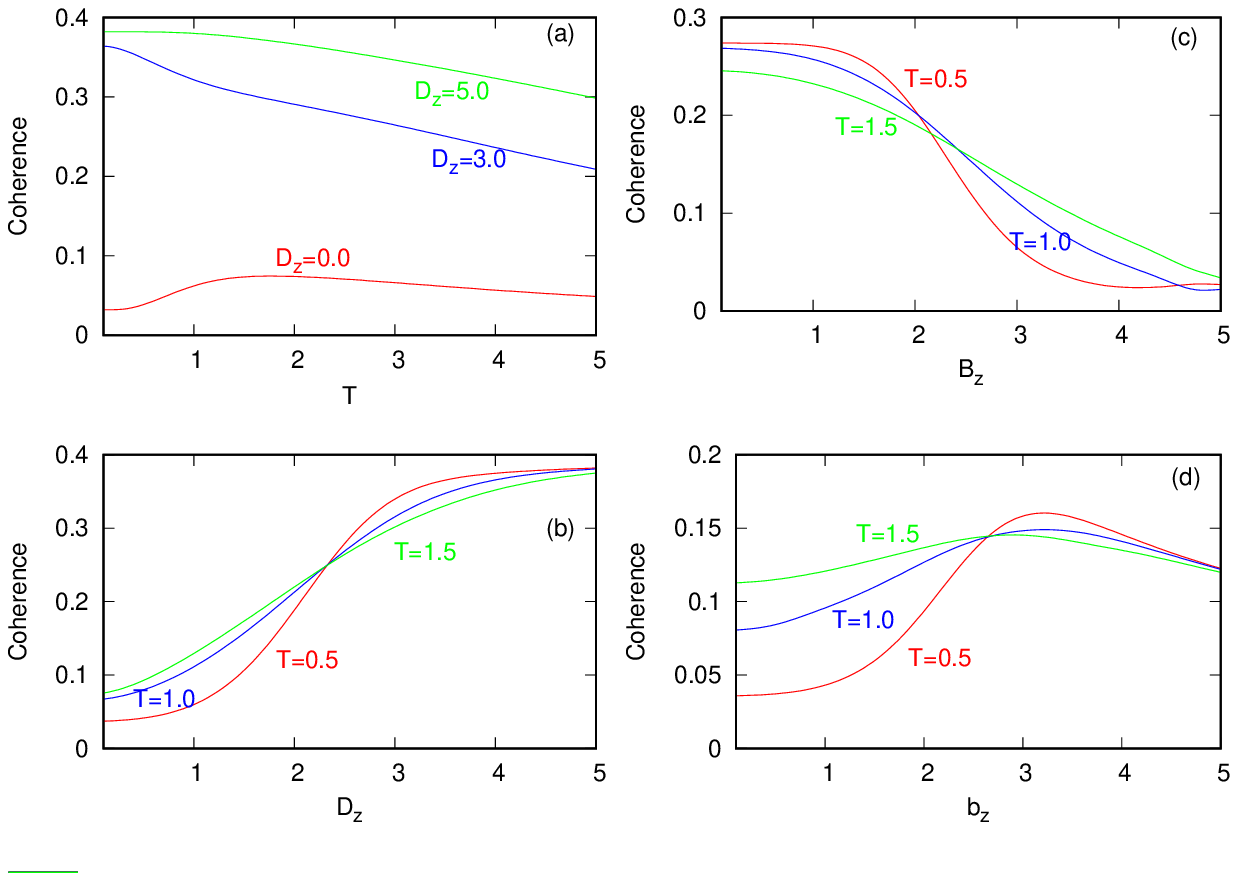}
\caption{Measurement of coherence in the $\sigma_{z}$ basis in the two-spin XYZ model with DM interactions in the $z$-direction and the external 
magnetic field in the $z$-direction.  (a) Quantum coherence versus temperature $T$ for different $D_{z}$, for $B_{z}=3.0$, $b_{z} =1.5$, 
$J_{x}=0.8$, $J_{y}=0.5$ and $J_{z}=0.2$.  (b) Quantum coherence versus $D_{z}$ for different temperatures $T$, for 
$B_{z}=3.0$, $b_{z} =1.5$, $J_{x}=0.8$, $J_{y}=0.5$ and $J_{z}=0.2$. (c) Quantum coherence versus $B_{z}$ for different temperatures $T$, 
for $D_{z}=1.0$, $b_{z} =1.5$, $J_{x}=0.8$, $J_{y}=0.5$ and $J_{z}=0.2$. (d) Quantum coherence versus $b_{z}$ for different temperatures 
$T$, for $D_{z}=1.0$, $B_{z}=3.0$,  $J_{x}=0.8$, $J_{y}=0.5$ and $J_{z}=0.2$.}
\label{fig2}
\end{figure}

The variation of quantum coherence as a function of the DM interaction and external field in the $\sigma_{z}$ basis
 is described through the plots in Fig. \ref{fig2}. The change of coherence with temperature is shown in Fig. \ref{fig2}(a) 
 for different values of the DM interaction parameter.  For larger values of the DM interaction parameter ($D_{x}=3.0,5.0$),  
 we find that the quantum coherence decreases with temperature. This 
is in line with the observation in Ref. \cite{li2008entanglement} and the well known effect of thermal decoherene on quantum systems.  
For $D_{z}=0$, the coherence decreases but there is an initial increase.  Through Fig. \ref{fig2}(b) we show the evolution of coherence with 
DM interaction.  Quantum coherence increases with $D_{z}$ and this occurs for all values of 
temperature.  The influence of magnetic field on quantum coherence is given in Figs. \ref{fig2}(c) and \ref{fig2}(d) for the 
homogeneous part and the inhomogenous part of the field respectively.  In Fig.  \ref{fig2}(c) we see that the  quantum coherence 
decreases with the average homogeneous magnetic field $B_{z}$.  From Fig. \ref{fig2}(d) we see that the coherence 
initially increases with the homogeneous field $b_{z}$, attains a maximum and then decreases to a saturation value.  

In summary, in this section we considered the situation where the DM interaction and the external magnetic field are along the same direction. 
In both the cases the quantum coherence shows a decrease with temperature. When we look into the variation of quantum coherence with the 
DM interaction parameter, the average homogeneous magnetic field $B$, and the inhomogenous magnetic field $b$ we find that the measurement basis has a 
definite outcome on the value of quantum coherence. Hence the qualitative behavior of quantum coherence is quite different for the 
situations described in Sec. \ref{samedirection} \ref{xdirectiondmandfield} and Sec. \ref{samedirection} \ref{zdirectiondmandfield} and 
this is a unique feature of quantum coherence which does not have an analog in entanglement measurements.

\section{DM interaction and the external field in different directions}
\label{differentdirection}
In this section we consider the two-site XYZ model in which the spin-lattice interaction and the external field are in different directions.  
Under these conditions we have three different cases as follows: ($i$) The DM interaction is along the measurement basis; 
($ii$) the external inhomogenous field is along the measurement basis and; ($iii$) both DM interaction and the external field 
are in different directions and orthogonal to the measurement basis.  The coherence of all these cases 
are examined in the discussion below.

\subsection{$\vec{D} \propto \vec{z}, \vec{B} \propto \vec{x} $ case} 
\label{dmzdirectionandfieldxdirection}
For the case with the DM interaction along the $z$-axis and the inhomogenous magnetic field along the 
$x$-axis, we have the Hamiltonian 
\begin{equation}
H = J_{x} \sigma_{1}^{x} \sigma_{2}^{x} + J_{y} \sigma_{1}^{y} \sigma_{2}^{y} 
    + J_{z} \sigma_{1}^{z} \sigma_{2}^{z}+D_{z}(\sigma_{1}^{x} \sigma_{2}^{y}-\sigma_{1}^{y} \sigma_{2}^{x})
    +(B_x + b_x ) \sigma_{1}^{x}+(B_x - b_x ) \sigma_{2}^{x}.
\label{xyz_ham_inhom_DzBx}     
\end{equation}
In the $\sigma^z$-basis the matrix form of the Hamiltonian is 
\begin{equation}
H = \left(
\begin{array}{cccc}
J_{z}&-b_{x}+B_{x}&b_{x}+B_{x}&J_{x}-J_{y}\\
-b_{x}+B_{x}&-J_{z}&2iD_{z}+J_{x}+J_{y}&b_{x}+B_{x}\\
b_{x}+B_{x}&-2iD_{z}+J_{x}+J_{y}&-J_{z}&-b_{x}+B_{x}\\
J_{x}-J_{y}&b_{x}+B_{x}&-b_{x}+B_{x}&J_{z}\\
\end{array}
\right).
\label{ham_mat_xyz_DzBx}
\end{equation}
From the Hamiltonian, one can calculate the density matrix $\rho(T)$ at thermal equilibrium.   Using the thermal density matrix
we can write down the diagonal form of the density matrix $\rho_{d}$ and using Eq. (\ref{qjsdmeasure}) we can calculate coherence 
in the system. 

\begin{figure}
\includegraphics[scale=1.25]{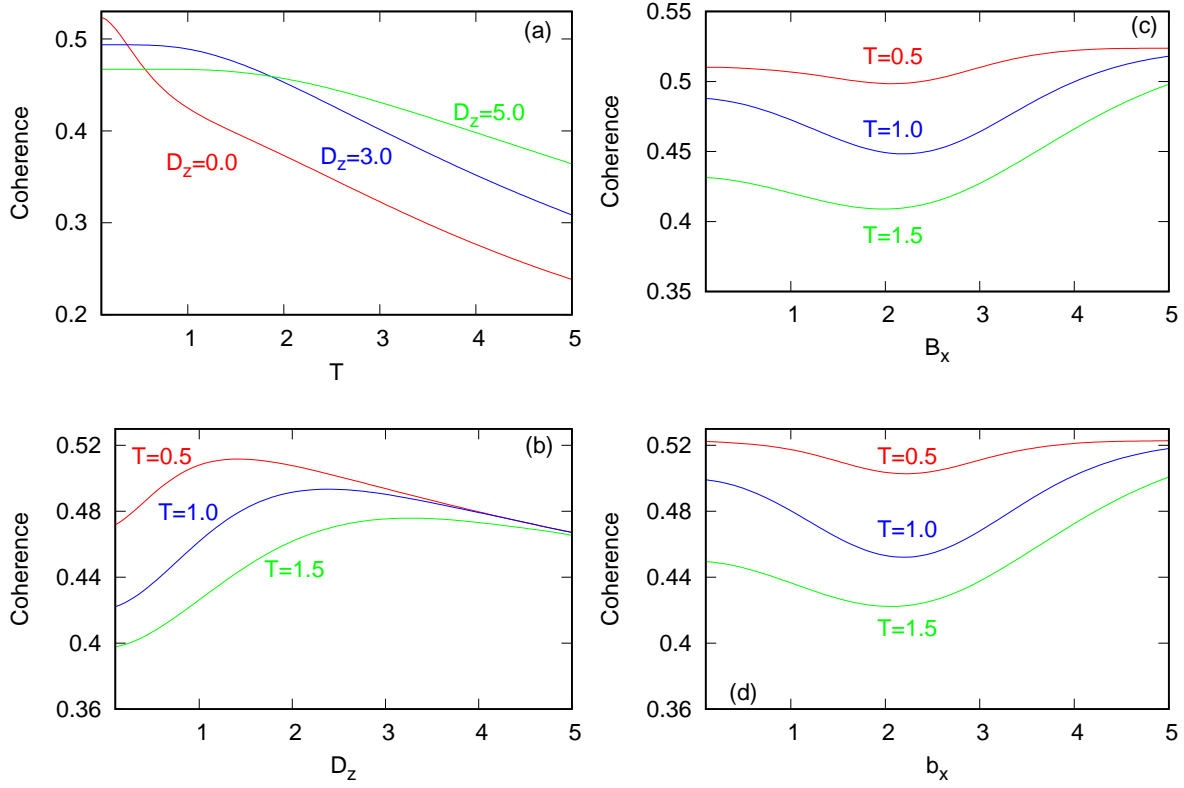}
\caption{The change in coherence measured in the $\sigma_{z}$ basis for the two-site XYZ model with DM interactions in the $z$-direction 
and the external field in the $x$-direction.  (a) Quantum coherence versus temperature $T$ for different $D_{z}$ for $B_{x}=3$, $b_{x} = 1.5$,
$J_{x}=0.8$, $J_{y}=0.5$ and $J_{z}=0.2$. (b)  Quantum coherence versus $D_{z}$, for different temperatures $T$, for $B_{x} =3.0$, 
$b_{x}=1.5$, $J_{x}=0.8$, $J_{y}=0.5$ and $J_{z}=0.2$.  (c) Quantum coherence versus $B_{x}$ for different temperatures $T$, for 
$D_{z}=1.0$,  $b_{x}=1.5$, $J_{x}=0.8$, $J_{y}=0.5$ and $J_{z}=0.2$. (d) Quantum coherence versus $b_{x}$ for different temperatures $T$, 
for $D_{z} =1.0$, $B_{x}=3.0$, $J_{x}=0.8$, $J_{y}=0.5$ and $J_{z}=0.2$.}
\label{fig3}
\end{figure}

For this model, the variation of quantum coherence with the different parameters is shown in Fig. \ref{fig3} with the 
measurement being carried out in the $\sigma_{z}$ basis. 
Thermal decoherence causes a loss of coherence due to increase in temperature and this is being observed in Fig. \ref{fig3}(a) for different
strengths of the DM interaction.  With respect to the DM interaction parameter, the coherence increases initially to attain a maximum 
value and then decreases.  This behavior is shown in Fig. \ref{fig3}(b) for different temperatures.  When the average homogeneous 
magnetic field $B_{z}$ is varied, the coherence decreases slightly to reach the minimum value and then increases back to a 
saturation value as shown in Fig. \ref{fig3}(c).  A similar behavior is observed for the variation of the inhomogenous 
component of the field $b_{z}$ as seen in Fig.  \ref{fig3}(d).  

\subsection{$\vec{D} \propto \vec{x}, \vec{B} \propto \vec{z}$ case}
\label{dmxdirectionandfieldzdirection}
When the magnetic field is oriented along the $z$-axis and the DM interaction is along the $x$-axis, the Hamiltonian
of the system reads:
\begin{equation}
H = J_{x} \sigma_{1}^{x} \sigma_{2}^{x} + J_{y} \sigma_{1}^{y} \sigma_{2}^{y} 
    + J_{z} \sigma_{1}^{z} \sigma_{2}^{z}+D_{x}(\sigma_{1}^{y} \sigma_{2}^{z}-\sigma_{1}^{z} \sigma_{2}^{y})
    +(B_z + b_z ) \sigma_{1}^{z}+(B_z - b_z ) \sigma_{2}^{z}.
\label{xyz_ham_inhom_DxBz}     
\end{equation}
The matrix form in the $\sigma^{z}$ basis is 
\begin{equation}
H = \left(
\begin{array}{cccc}
2B_{z}+J_{z}&iD_x&-iD_x&J_{x}-J_{y}\\
-iD_x&2b_{z}-J_{z}&J_{x}+J_{y}&iD_x\\
iD_x&J_{x}+J_{y}&-2b_{z}-J_{z}&-iD_x\\
J_{x}-J_{y}&-iD_x&iD_x&-2B_{z}+J_{z}\\
\end{array}
\right).
\label{ham_mat_xyz_inhom_DxBz}
\end{equation}
The thermal density matrix $\rho(T)$ can be calculated from the Hamiltonian (\ref{ham_mat_xyz_inhom_DxBz}) using
which we can find the quantum coherence of the system from Eq. (\ref{qjsdmeasure}).  

\begin{figure}
\includegraphics[scale=1.25]{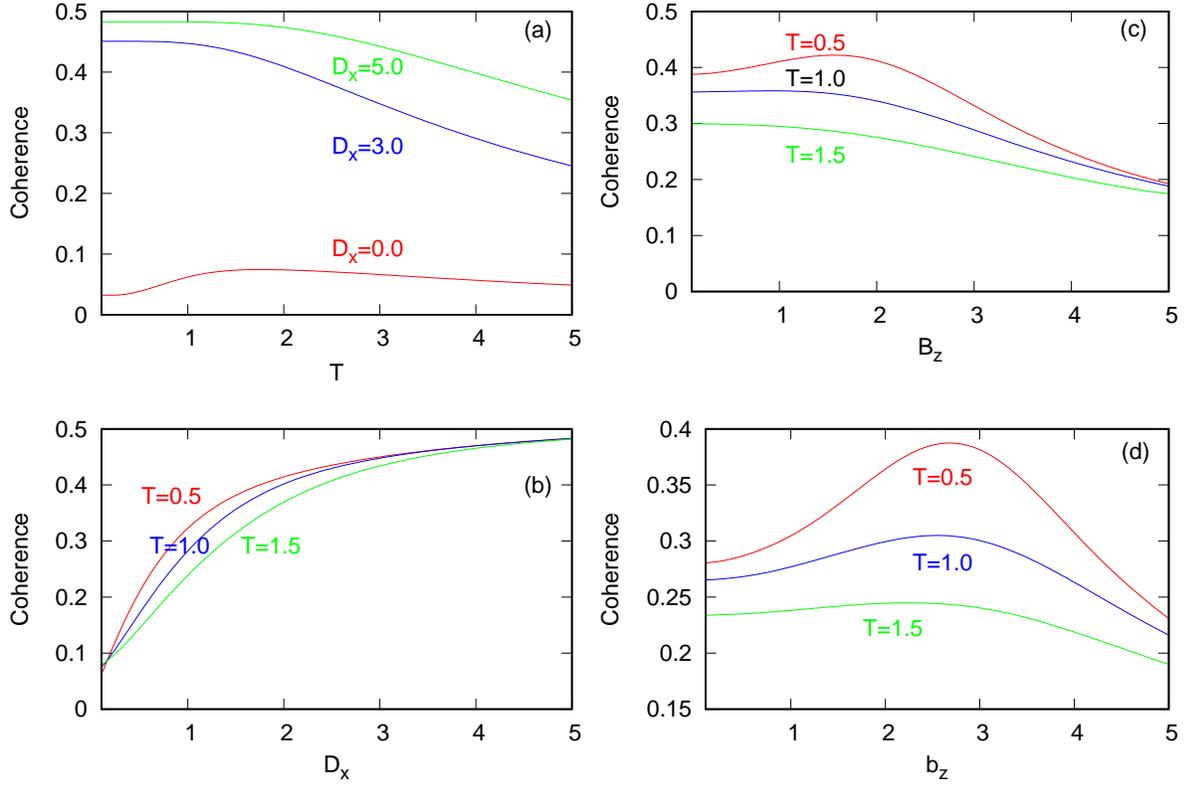}
\caption{The variation in coherence is measured in the $\sigma_{z}$ basis for the two-spin XYZ model with DM interactions in the $x$-direction 
and the external magnetic field in the $z$-direction.  (a) Quantum coherence versus temperatures $T$ for different $D_{x}$ for $B_{z}=3$, $b_{z} = 1.5$,
$J_{x}=0.8$, $J_{y}=0.5$ and $J_{z}=0.2$. (b) Quantum coherence versus $D_{x}$, for different temperatures $T$, for $B_{z} =3.0$, 
$b_{z}=1.5$, $J_{x}=0.8$, $J_{y}=0.5$ and $J_{z}=0.2$. (c) Quantum coherence versus $B_{z}$ for different temperatures $T$, for 
$D_{x}=1.0$,  $b_{z}=1.5$, $J_{x}=0.8$, $J_{y}=0.5$ and $J_{z}=0.2$. (d) Quantum coherence versus $b_{z}$ for different temperatures $T$, 
for $D_{x} =1.0$, $B_{z}=3.0$, $J_{x}=0.8$, $J_{y}=0.5$ and $J_{z}=0.2$.}
\label{fig4}
\end{figure}

The dependence of quantum coherence with temperature 
is shown in Fig. \ref{fig4}(a).  As before, we find that the coherence decreases with temperature as expected due to thermal effects removing coherence.   With increase in the value of the DM interaction 
parameter, we find that quantum coherence increases monotonically as seen in Fig \ref{fig4}(b).  From Fig \ref{fig4}(c)
we can see that coherence decreases with the average homogeneous field $B_{z}$.  While this is uniform for higher 
temperatures, for lower values of temperature the coherence increases to a peak value and then decreases.  The same 
behavior is also replicated when we look into the variation of coherence with the inhomogenous field $b_{z}$ as seen in Fig. \ref{fig4}(d).   

\subsection{$\vec{D} \propto \vec{x}, \vec{B} \propto \vec{y} $ case}
\label{dmxdirectionandfieldydirection}
\begin{figure}
\includegraphics[scale=1.25]{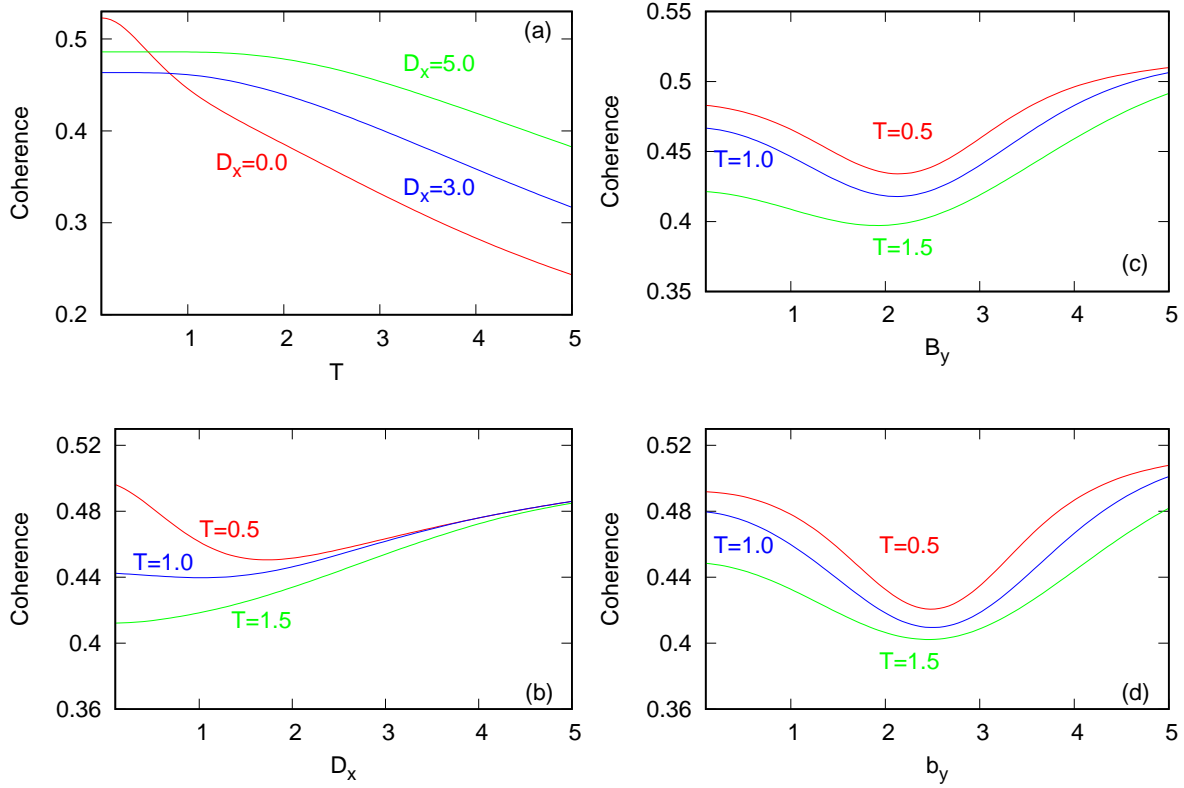}
\caption{The coherence variation measured in the $\sigma_{z}$ basis for the XYZ model with DM interactions in the $x$-direction 
and the external field in the $y$-direction.  (a) Quantum coherence versus temperature $T$ for different $D_{x}$ for $B_{y}=3$, $b_{y} = 1.5$,
$J_{x}=0.8$, $J_{y}=0.5$ and $J_{z}=0.2$.  (b) Quantum coherence versus $D_{x}$, for different temperatures $T$, for $B_{y} =3.0$, 
$b_{y}=1.5$, $J_{x}=0.8$, $J_{y}=0.5$ and $J_{z}=0.2$. (c) Quantum coherence versus $B_{y}$ for different temperatures $T$, for 
$D_{x}=1.0$,  $b_{y}=1.5$, $J_{x}=0.8$, $J_{y}=0.5$ and $J_{z}=0.2$. (d) Quantum coherence versus $b_{y}$ for different temperatures $T$, 
for $D_{x} =1.0$, $B_{y}=3.0$, $J_{x}=0.8$, $J_{y}=0.5$ and $J_{z}=0.2$. }
\label{fig5}
\end{figure}

An interesting situation arises when the DM interaction and the external field are perpendicular to each other and neither of them 
are oriented along the measurement basis.  The Hamiltonian in this case is
\begin{equation}
H = J_{x} \sigma_{1}^{x} \sigma_{2}^{x} + J_{y} \sigma_{1}^{y} \sigma_{2}^{y} 
    + J_{z} \sigma_{1}^{z} \sigma_{2}^{z}+D_{x}(\sigma_{1}^{y} \sigma_{2}^{z}-\sigma_{1}^{z} \sigma_{2}^{y})
    +(B_y + b_y ) \sigma_{1}^{y}+(B_y - b_y ) \sigma_{2}^{y}.
\label{xyz_ham_inhom_DxBy}     
\end{equation}
From the knowledge of the Hamiltonian we can calculate $\rho(T)$ the thermal density matrix of the system and calculate the 
coherence using Eq. (\ref{qjsdmeasure}).  Our results are shown in Figs. \ref{fig5}(a)-(d).  
  
The quantum coherence decreases with temperature 
as seen in Fig. \ref{fig5}(a), as before due to thermal decoherence. The qualitative behavior of coherence with respect to change in the DM interaction parameter, 
the external homogeneous field $B_{z}$, and the inhomogeneous field $b_{z}$ is shown in Figs. \ref{fig5}(b), (c) and (d) respectively.  The presence of spin-lattice interaction gives rise to new or enhanced contributions to off-diagonal elements and hence the overall effect is to 
increase the quantum coherence in the system.  However in the XYZ model with both DM interaction and field in the $x$-direction, there is initially
a decrease in coherence. After reaching the minimum the coherence again starts to increase.  This dip occurs because the coherence arises due to 
two different interactions namely the spin-spin and the spin-lattice interaction.  While the amount of coherence due to the spin-spin interaction 
decreases, the coherence due to the spin-lattice increases with the DM parameter $D_{x}$.  There is a mismatch between the two rates of 
change leading to the formation of a dip in the amount of coherence.  For higher values of temperature no dip is observed.  In the case of the 
other two cases described in Sec. \ref{differentdirection} \ref{dmzdirectionandfieldxdirection} and 
Sec. \ref{differentdirection} \ref{dmxdirectionandfieldydirection}, the coherence increases with  $D_{x}$ for higher values of temperature.  
For lower values of temperature, the coherence decreases with DM parameter.  This special behavior is also due to the competing rates 
of change of coherence of the spin-spin and spin-lattice interactions.  Hence, overall we can conclude that coherence increases with the 
DM interaction, even if the effect may not be so easily observed.  

In summary, when an external field is applied on a spin system, the field changes the quantum properties by influencing the spins.  In our work, 
we study the effect of an inhomogenous external field.  The external field can be divided into an homogeneous component $B$ 
and an inhomogenous component $b$.  The strength and the direction of the homogeneous component $B$ of the field is the same
for the spins at both the lattice sites.  In the case of the homogeneous component $b$, the strength of the field acting on both 
the spins is the same, but their direction is different.  From Figs. \ref{fig1} - \ref{fig5} we observe that the quantum coherence 
decreases when the direction of the homogeneous field $B$ is the same as the measurement basis. On the other hand, the 
coherence increases when the measurement basis and the external field are orthogonal to each other.  In the case of the 
inhomogenous component of the field, the coherence mostly increases except for the isolated case of XYZ model with 
DM interaction along the $x$-direction and the field $b$ is along the $z$-direction.

%
%
%
%

\section{Coherence variation under two control parameters}
\label{two control parameters}

Up to this point we have not considered the dependence of the spin-spin interaction $J_{i}$, which has been considered a constant.  In this section, we examine the variation of 
coherence with respect to the simultaneous variation of the spin-spin  interaction parameter and one of the other control parameters such as the DM parameter or the external field.  

\begin{figure}
\includegraphics[scale=1.25]{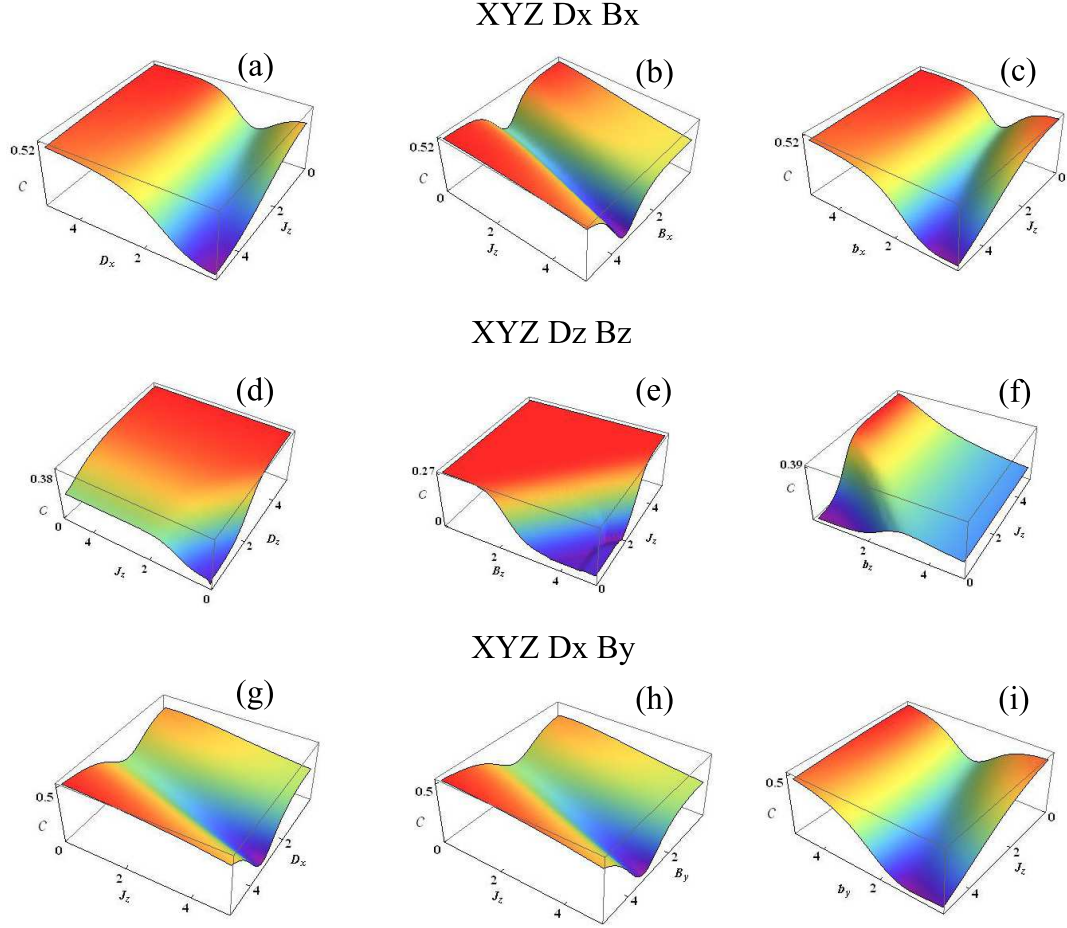}
\caption{The variation of coherence in the XYZ model with DM interaction and external field along the $x$-axis is shown through the first row 
of plots as follows: (a) coherence Vs  $J_{z}$ Vs $D_{x}$  (b) coherence Vs $J_{z}$ Vs $B_{x}$ (c) coherence Vs $J_{z}$ Vs $b_{x}$.  
The fixed values are $D_{x} = 1.0$, $B_{x} = 3.0$, $b_{x} =1.5$, $J_{x} =0.8$, $J_{y} = 0.5$ and $T=0.5$.  In the second row, the XYZ model with
DM interaction and the field in the $z$-direction is given as follows: (d) coherence Vs  $J_{z}$ Vs $D_{z}$  
(e) coherence Vs $J_{z}$ Vs $B_{z}$ (f) coherence Vs $J_{z}$ Vs $b_{z}$.  The fixed values are 
$D_{z} = 1.0$, $B_{z} = 3.0$, $b_{z} =1.5$, $J_{x} =0.8$, $J_{y} = 0.5$ and $T=0.5$.  The final row describes the XYZ model with 
DM interaction in the $x$-direction and the field along the $y$-direction as follows: (d) coherence Vs  $J_{z}$ Vs $D_{x}$  
(e) coherence Vs $J_{z}$ Vs $B_{y}$ (f) coherence Vs $J_{z}$ Vs $b_{y}$.  The fixed values are 
$D_{x} = 1.0$, $B_{y} = 3.0$, $b_{y} =1.5$, $J_{x} =0.8$, $J_{y} = 0.5$ and $T=0.5$. All the measurements are made uniformly in 
the $\sigma_{z}$ basis.}
\label{fig6}
\end{figure}

The quantum coherence in the two-spin XYZ model with DM interaction and the external field along the $x$-direction is 
described in the first row of plots in Figs. \ref{fig6} (a)-(c).  Fig. \ref{fig6}(a) shows the variation of coherence 
with respect to the spin-spin interaction parameter $J_{z}$ and the DM interaction parameter $D_{x}$.  From the plot 
we observe that for a given a constant temperature, the amount of quantum coherence decreases with the 
spin-spin interaction parameter $J_{z}$. Contrarily on increasing $D_{x}$ the amount of coherence increases.  This 
implies that the spin-spin interaction and the spin-lattice interaction have counteracting effects on the amount of coherence 
in the system.  Fig. \ref{fig6}(b) shows the quantum coherence variation with respect to the parameters $J_{z}$ 
and $B_{x}$ (homogeneous component of the external field).  Again we find that the coherence decreases with $J_{z}$, 
but increases with the external field.  The external field $B_{x}$ tries to align the spins in a uniform direction and hence we 
observe an increase in the amount of quantum coherence.  Fig. \ref{fig6}(c) shows coherence 
as a function of the spin-spin interaction $J_{z}$ and the inhomogenous component of the field $b_{x}$.  This plot 
exhibits the features where we find that coherence decreases with $J_{z}$, whereas it increases with $b_{x}$.  
These observations are identical to the results obtained in  Fig. \ref{fig6}(a) and \ref{fig6}(b).

In Figs.  \ref{fig6} (d)-(f) the quantum coherence features of the two-site anisotropic model with DM interaction and 
the field in the $z$-direction is shown.  We choose the measurement basis also to be in the $\sigma_{z}$ basis 
and consider a constant temperature $T$.  From Fig. \ref{fig6}(d) we find that the coherence increases with $J_{z}$  
as well as $D_{z}$.  In this situation, there is no counteracting influence between the spin-spin and 
spin-lattice interactions.  Rather they exhibit a co-operative nature and enhances the coherence in the system with 
increase in these parameters.  In Fig. \ref{fig6}(e) we notice that coherence decreases with increase in $B_{z}$ and 
it increases with $J_{z}$.  Finally in Fig. \ref{fig6}(f) we observe that coherence increases with the inhomogenity of the 
external field $b_{z}$ as well with the increase of the spin-spin interaction parameter $J_{z}$.  

Finally, in Figs. \ref{fig6} (g)-(i) we look into the coherence dependence of the model with the spin lattice interaction in 
the $x$-direction and measurement along the $\sigma_{z}$ basis.  The external inhomogenous field is along the 
$y$-direction.  When the spin-spin interaction $J$ is increased, the coherence decreases. But when the spin lattice 
interaction is increased, the coherence increases.  These results are shown in Fig. \ref{fig6}(g).  Also through the 
plots Fig. \ref{fig6}(h) and Fig. \ref{fig6}(i) we find that coherence decreases with increase in the spin-spin 
interaction $J_{z}$.  On increasing the homogeneous component $B_{y}$ and inhomogenous component 
$b_{y}$ we observe that the coherence increases.  

In Table \ref{table1}, based on the two dimensional plots, we summarize the variation of quantum coherence 
with different parameters for all the models.  Using this in conjunction with  Fig. \ref{fig6} we 
arrive at the following conclusions. We find that when the  direction of the DM interaction and external field is orthogonal to the 
measurement, the coherence in the system decreases with the spin-spin coupling $J$ value.  Similarly,
the coherence increases with the spin-lattice interaction parameter $D$.  Also in this situation both the 
homogeneous and inhomogenous fields helps to increase the coherence in the system.  Under the conditions when 
the direction of measurement basis is same as the direction of the spin-lattice, the spin-spin interaction and the 
DM interaction increases the quantum coherence in the system. With respect to the external field, the coherence
decreases with increase in the homogeneous component of the field $B$ and it increases with the inhomogenous
component of the field $b$.

\begin{table}
\caption{The variation of the coherences for the different parameters is summarized below. Here $+$ means coherence increases with 
the variable and $-$ means coherence decrease with the variable. The measurement basis is always in the $ z$-direction.  }
\begin{center}
\begin{tabular}{|p{3cm}|p{2.2cm}|p{2.2cm}|p{2.2cm}|p{2.2cm}|p{2.2cm}|}
\hline
Parameters & $\vec{D} \propto \vec{x}$ & $\vec{D} \propto \vec{z}$ & $\vec{D} \propto \vec{z}$ & $\vec{D} \propto \vec{x}$ & $\vec{D} \propto \vec{x}$  \\
                    & $\vec{B} \propto \vec{x}$  & $\vec{B} \propto \vec{z}$ & $\vec{B} \propto \vec{x}$ & $\vec{B} \propto \vec{z}$ & $\vec{B} \propto \vec{y}$   \\
\hline 
Temperature $T$ & $-$ & $-$ & $-$ & $-$ & $-$\\
\hline 
DM interaction $D$ & $+$ & $+$ & $+$ & $+$ & $+$\\
\hline
Homogeneous magnetic field $B$ & $+$& $-$ & $+$ & $-$ & $+$\\
\hline
Homogeneous magnetic field $b$ & $+$ & $+$ & $+$ & $-$ & $+$\\
\hline 
\end{tabular}
\label{table1}
\end{center}
\end{table}

%
%
%
\setcounter{equation}{0}
\section{Conclusions}
\label{conclusions}

The finite temperature quantum coherence of the two-spin XYZ Heisenberg model with Dzyaloshinsky-Moriya (DM) interaction subjected to an 
external inhomogenous magnetic field was investigated. In the investigation of these two-site models we consider two major cases 
namely: ($i$) the DM interaction and the external magnetic field are in the same direction;  
($ii$) when the DM interaction and the external field are not in the same direction.  
Since quantum coherence is a basis-dependent quantity, these two classes are further divided 
based on the relation between the measurement basis and the direction of the DM parameter and the field. 
We first confirm that quantum coherence always decreases with temperature, due to the commonly observed effect that quantum 
coherence is destroyed by thermal decoherence.   Next we find that the 
quantum coherence always increases with the strength of the DM interaction parameter irrespective of the relation 
between the spin-lattice interaction and the measurement basis.   This result is similar to the one obtained for 
entanglement measured using concurrence.   The quantum coherence decreases when the direction of the measurement basis is the 
same as that of the spin-spin interaction.  A similar behavior of quantum coherence is observed when the external field is varied.  

A spin chain always contains a spin-spin interaction apart from the DM interaction and the external field as 
described in our work.  We investigated the change in coherence when the spin-spin interaction is simultaneously 
varied with the DM interaction parameter or the external fields.  From the observations we find that the 
spin-spin interaction and the spin-lattice interaction have counteracting effects on the quantum coherence 
of the system when the direction of the measurement is orthogonal to the direction of the spin-lattice 
interaction.  We note that this study could be further extended to larger systems, and studying spin models with 
staggered DM interactions \cite{ma2011quantum,miyahara2007uniform} and random field interactions
\cite{bruinsma1983one,fujinaga2007entanglement}.  

The amount and distribution of quantum coherence is connected to quantum information processing tasks like 
dense coding and quantum teleportation \cite{pan2017complementarity}.  For example, the optimal dense coding capacity of a bipartite system
is inversely related to the amount of local coherence in the system \cite{pan2017complementarity}.  Similarly, there is an inverse relation between the teleportation fidelity
and the local coherence in a system \cite{pan2017complementarity}.  From our earlier studies \cite{radhakrishnan2016distribution} we know that coherence can 
exist localized within a qubit or as correlations between the qubits.  Hence a small amount of local coherence must imply a higher amount 
of global coherence, which naturally means strongly correlated quantum systems help in tasks such as dense coding and quantum teleportation.
In the case of the spin qubits considered, we observe that the total coherence in the system is always global coherence which 
implies that all the available form of coherence in a spin qubit can be used for quantum information processing tasks.  In addition,
the existence of several control parameters such as the spin-spin interaction parameter, the spin-lattice interaction parameter and the external 
parameters  such as magnetic field and temperature helps us to modify the state of a system to improve the  capacity of dense coding 
and the teleportation fidelity of quantum states.  In Ref. \cite{pan2017complementarity} the authors only demonstrated the complementarity 
relations of the local coherence with the dense coding capacity and the teleportation fidelity.  An important development in this 
direction would be to directly quantify the dense coding capacity and teleportation fidelity in terms of the global coherence.  Such an 
improvement will be directly applicable to the spin systems we have considered in our work, where we can quantify the ability to 
perform quantum information processing tasks using the coherence of the spin system.

%
%
%
\section*{Acknowledgments}
TB and RC are supported by the Shanghai Research Challenge Fund; New York University Global Seed Grants for Collaborative Research; 
National Natural Science Foundation of China (61571301,D1210036A); the NSFC Research Fund for International Young Scientists 
(11650110425,11850410426); NYU-ECNU Institute of Physics at NYU Shanghai; the Science and Technology Commission of Shanghai 
Municipality (17ZR1443600); the China Science and Technology Exchange Center (NGA-16-001); and the NSFC-RFBR Collaborative 
grant (81811530112).

%
%
%


\end{document}